\documentstyle [12pt,setspace,dina4] {article}

\title{Electromagnetic Form Factor of the Nucleon in the Time-like
       Region\thanks{Work supported by GSI Darmstadt and BMFT}}
\author{M.  Sch\"afer\thanks{Part of dissertation of M. Sch\"afer},
        H. C. D\"onges and U. Mosel\\
        Institut fuer Theoretische Physik, Universit\"at Giessen\\
        D--35392 Giessen, Germany}

\begin{document}
\pagestyle{empty}

\titlepage

\maketitle
\begin{abstract}
\par
\begin{center} submitted to Physics Letter B
\end{center}
\par
We  explore the possibility to verify vector meson dominance
for   the  nucleon  by  measuring  the  half-off-shell  form factor  in  the
vector-meson  mass  region.  Cross  sections  for  the  process  $\gamma  p
\rightarrow p + e^+e^-$ be presented.
\end{abstract}

\newpage
\pagestyle{plain}

There exists by now a vast amount of experimental data on the
electromagnetic form factor of the nucleon \cite{Hoehler1,Dubnicka}.
Most of these data were
obtained with electron scattering experiments yielding information on the
form factor in the space-like sector. The form factor in the time-like
region has been explored by $p \bar{p}$ annihilation experiments; because
of the masses of the particles involved this channel is open only for
$q^2 > 4{m_p}^2 \approx 3.6 \ {\rm GeV}^2$.

To explain the coupling of photons to hadrons vector meson dominance (VMD) has
been invoked; in this picture one assumes that the photons couple to
hadrons only through the vector mesons. While this picture is well
established for the pion,  the situation for the nucleon is less
obvious \cite{Jaffe}. The experimentally determined form factors have been
shown to follow very well the so-called dipole fit \cite{Perkins}
\begin{equation}
G_E(q) =  \left(  \frac{m_v^2}{q^2 -  m_v^2}  \right)^2
\end{equation}
where $m_v^2 \approx 0.71 ~GeV^2$. The fact that not a monopole,  but a dipole
dependence appears here presents a problem for naive vector meson
dominance since the form factor should contain the vector meson propagator
only linearly. Furthermore,  coupling constants often agree only within
factors of 2 with the predictions of vector meson dominance \cite{Rho,Neil}.

There are essentially two classes of explanations for the unexpected
$q$-dependence of the propagators. In the first one assumes that the
observed form factor is the {\em sum} of contributions of several vector
mesons and that these contributions conspire such that the
effective $q$-dependence is of dipole type \cite{Hoehler1}. To this class
of models we also count the attempts to describe the form factor as a sum
of two terms,  one describing the coupling of the photon to the meson
cloud (vector meson dominance) and a second giving the direct coupling of
the photon to the quark core \cite{Rho}. In the second class of
explanations one assumes that the effective form factor is the {\em
product} of the vector meson propagator and a cut-off form factor of
monopole type that takes the properties of perturbative QCD at large
$q^2$ into account \cite{Gari}. Both of these models,  if
extended into the timelike sector,  give large,  resonance-like
contributions at the vector-meson mass $m_v \approx 0.78$ GeV
\cite{Mosel} which,  of course,  have never directly been seen because
they lie in
the so-called unphysical region which is not accessible in on-shell
processes. There the models as well as the dispersion theoretical analyses
\cite{Hoehler2} have to be considered as analytic continuations from
the space-like measurements into the time-like region, and, more
specifically, to the on-shell point there.
The direct experimental proof for VMD for the nucleon,  the presence of
a clear resonance
in its electromagnetic form factor just as for the pion,  is
thus still missing.

In this letter we explore the possibility to ascertain the resonance
behavior of the form factor expected from VMD by determining
experimentally the {\em half-off-shell} form factor of the nucleon in the
time-like region just mentioned. If measured,  this form factor could then be
analytically continued to the on-shell point even at time-like $q$ in the
unphysical region. This continuation from a time-like half-off-shell point
may supplement and be even more conclusive and more
stringent than the continuation from spacelike, on-shell points.

One way to explore the half-off-shell form factor in the time-like
region is that of dilepton production in elementary processes. The DLS
group at the BEVALAC \cite{Roche} has recently undertaken such
experiments both for $p + p$ and $p + d$ reactions.
First theoretical analyses \cite{Schaefer,Kaempfer} indicate a
sensitivity of the invariant-mass spectra to the electromagnetic
form factor. However, the results of these
experiments do not allow for an easy,  unambigous extraction
of the form factor,  because the strong-interaction vertices affect the
cross-section. The same is true for quite early experiments employing the
reaction $p + \pi \rightarrow p + e^+e^-$ \cite{p-pi}.

In order to avoid the ambiguities necessarily connected with the presence
of strong-interaction vertices and form factors we propose to explore a
purely electromagnetic reaction. The process we consider is
Compton-scattering of a real photon into the time-like sector,  i. e. the
process $\gamma p \rightarrow p + e^+e^-$.

Real Compton scatttering in the energy range considered here, i.e. up to
1 GeV photon energy, has been investigated in refs. \cite{Ishii}
and \cite{Wada}.
These authors have given a decomposition of
the Compton scattering amplitude
\begin{equation}          \label{amplitude1}
A = A^R + A^B \exp(-C (1 - \cos \vartheta) )
\end{equation}
into a resonance contribution $A^R$ and into a non-resonant Born amplitude
$A^B$.

The Born amplitudes are given in detail in ref. \cite{Ishii}, the
exponential multiplying it damps the backward-angle scattering and
represents an ad-hoc correction that takes neglected effects such
as the $t$-channel scattering into account; $\vartheta$ is the photon
scattering angle in the c.m. frame.

While for the resonance amplitude the {\it u}-channel is neglected,
the only contribution to this amplitude is the {\it s}-channel \cite{Wada}
\begin{eqnarray}          \label{amplitude2}
A_{\lambda \mu} = \frac{k_0}{k}
   \frac{2 \gamma^\lambda \gamma^\mu W} {M_r^2 - W^2 - iW\Gamma_r}
   \quad e^{i\delta_r}
\end{eqnarray}
with
\begin{eqnarray*}
k^2(q^2,W^2) &=& \left( \frac{W^2-m_N^2+q^2}{2W} \right)^2 - q^2 \nonumber \\
k_0^2 &=& k^2(q^2=0,W^2=M_R^2) ~.
\end{eqnarray*}
Here $M_r$ is the
mass of the resonance, $\Gamma_r$ its width and W the invariant energy.
The phases $\delta_r$ are introduced to take the relative
phases between the various resonances and the Born amplitude into account.
For later purposes we have written eq.(\ref{amplitude2}) in a form
also correct for off-shell photons ($q^2\neq 0$).
Essential for our purposes here is the structure of the elastic photon
width $\gamma^\lambda$ in eq. (\ref{amplitude2}) which is given by
\begin{eqnarray}              \label{photonwidth}
\gamma^\lambda = \gamma_0^\lambda \left( \frac{k}{k_0} \right)^{j_0}
                   F_\gamma(q^2=0,W^2)
\end{eqnarray}
with the form factor
\begin{eqnarray*}
F_\gamma(q^2,W^2)=
    \left( \frac{k_0^2 + X^2}{k^2 + X^2} \right)^{j_0/2} ~ .
\end{eqnarray*}
Here $j_0$ is the spin of the resonance and $\gamma_0^\lambda$ the
photon coupling constants of the resonance.
For all further details see Wada et al.
\cite{Wada}. The analysis
includes all resonances up to a mass of about 2 GeV;
in the following we use the parameters extracted there from fitting
Compton scattering data in the energy range up to 1.2 GeV.

In order to use this information on the electromagnetic vertices of the
nucleon for our purpose of investigating the electromagnetic form factor
of the nucleon in the time-like sector, we now
consider the general
structure of the electromagnetic vertex appearing in the $\gamma + p
\rightarrow p + e^+e^-$ process. We first note that both electromagnetic
vertices are half-off shell, at the entrance point for an on-shell
photon and at the exit point for a virtual time-like photon. The general
structure
of the electromagnetic half-off-shell vertex has recently been investigated
in detail by Tiemeijer and Tjon \cite{Tiemeijer-Tjon} and Naus and Koch
\cite{Naus-Koch}. These authors have shown that the general
electromagnetic form factor for a half-off-shell vertex depends on the
squares of the fourmomentum of the photon, $q^2$,  and of the off-shell
nucleon, $W^2$,
$F = F(q^2,W^2)$. The reduction of the original 6 different
form factors to only 1 is possible when using
gauge-invariance, particle-antiparticle symmetry \cite{Gross-Riska} and
the experimentally well established equality of electric and magnetic
form factors in the space-like region.

In order to explore the sensitivity of the dilepton invariant
mass spectra on
a possible VMD form of the form factor
we now extend the electromagnetic form factor (\ref{photonwidth}) into
the region $q^2\neq 0$ and multiply a VMD based factor
$F_{VMD}$ to it.
We thus make the ansatz
\begin{equation}   \label{formfact2}
F(q^2,W^2) = F_{VMD}(q^2) F_\gamma(q^2,W^2)
\end{equation}
with
\begin{eqnarray*}
F_{VMD}(q^2) = \frac{m_V^2}{m_V^2 - q^2}
\end{eqnarray*}
for the form factor in the timelike region. The singularity at
$q^2=m_V^2$ is avoided by implementing the decay width of
the vectormesons in the denominator of $F_{VMD}$ in (\ref{formfact2}).
For simplicity we assume the same form factor for all the hadrons
in the timelike region. This is supported by the analysis of Stoler
that shows that the spacelike form factors of all the resonances
agree within a few percent for $q^2 > -1 ~ GeV^2$ \cite{Stoler}.

It is now straigthforward to generalize the amplitude-based description
of Compton scattering of ref. \cite{Wada} to virtual photons. This is
done by using in eqs. (\ref{amplitude2}) and (\ref{photonwidth})
the momentum of a massive photon created in the resonance decays.
In addition to the real Compton amplitude the
photon vertex $\gamma^\lambda$ is replaced
\begin{eqnarray}
\gamma^{(\lambda=\lambda_f-\mu_f)} =
               &J^{(\mu_f)} \cdot \epsilon^{*(\lambda_f)}&
               \Rightarrow       \nonumber \\
  \Bigg\{
          &\left( J^{(\mu_f)} \cdot \epsilon^{*(+1)} \right)&
          \left( J^{e^+e^-} \cdot \epsilon^{(+1)} \right)  \nonumber \\
+         &\left( J^{(\mu_f)} \cdot \epsilon^{*(-1)} \right)&
          \left( J^{e^+e^-} \cdot \epsilon^{(-1)} \right)  \nonumber \\
+         &\left( J^{(\mu_f)} \cdot \epsilon^{*(0)} \right)&
          \left( J^{e^+e^-} \cdot \epsilon^{(0)} \right)
          \frac{M^2}{k_o^2}
  \Bigg\} \frac{1}{M^2} ~;
\end{eqnarray}
the superscripts in parenthesis give the helicity quantum numbers,
$M$ is the invariant mass of the dilepton pair,  $1/M^2$ the
massive photon's propagator and $J^{e^+e^-}$ the dilepton current.

In this way we take into account that the massive photon
can be polarized in longitudinal direction.
For the longitudinal contribution we assume
\begin{equation}
J^{(\mu_f)} \cdot \epsilon^{*(0)} = \gamma^{(-\mu_f)} ~;
\end{equation}
thus we can use for this amplitude also the photon coupling constant
$A_{1/2}$ from ref.\cite{Wada}.
The contribution from the nucleon poles is calculated exactly.
In this way we retain the experimentally determined half-off-shell
vertex for real photons and extend it to the time-like photon
vertex. We also ensure by construction that the cross-section for
dilepton production reduces at the photon point to the experimentally
correct Compton scattering value.

Within this description we are able to calculate the amplitude for the
Feynman diagrams shown in Figs. 1a and 1b. For the creation
of virtual photons we also have to add contributions to the
amplitude from the so called {\it Bethe-Heitler} diagrams (Fig. 1c);
for the corresponding space-like form factor we use the measured one.
Because of the pole-like behaviour of the electron propagator
this contribution dominates in any integrated cross section.
Therefore, we look only at a five-times differential cross section,
very similiar to what was done in \cite{WALECKA}. In this symmetric
kinematical situation (Fig. 2) it was shown that the influence of the
Bethe-Heitler diagrams decreases with increasing angle $\theta$
(ref.\cite{WALECKA}).
The invariant mass of the dilepton pair is a simple function
of only this opening angle.

It is possible in this kinematical situation that a given invariant mass
corresponds to two different angles, one less and one larger than $90^o$.
For $\theta < 90^o$ we show only the invariant mass spectra above
400 MeV, because for smaller masses (=smaller angles) the contribution
from the Bethe-Heitler diagrams dominates the spectrum. Even
for larger invariant masses the Bethe-Heitler contribution is the
main contribution to the cross section (Fig. 3) if we switch
off the VMD form factor for the hadrons by setting $F_{VMD}=1$.
Since the contribution of the nucleons
(Fig. 3: dotted line) is dominated by the Bethe-Heitler diagrams,
it shows only a weak resonance shape. But a resonance behaviour
is definitely seen for the
contribution of the nucleon resonances
which is, now including
the VMD, larger than the Bethe-Heitler contribution. Because
of the weak sensitivity of the cross section to the nucleon
form factor it may be possible to get information on the
electromagnetic form factor for the resonances if the detection angle
of the leptons is less than 90 degree, but the signal sits on a
steeply falling background.

If the angle is larger than $90^0$ (Fig. 4), we find a strong effect
of the VMD on the nucleon as well as the resonance contribution,
because now the influence of the Bethe-Heitler diagrams is negligible.
For this kinematic situation
we see in Fig. 4 that the resonance contribution
is now larger than the nucleonic one by roughly a factor of two.

After fixing the resonance form factors $F^R$ in the forward region it
might be possible to extract from the
cross section in backward kinematics the form factors
$F^N_i$ for the nucleon.
Because the resonances are propagated with $q_R^2 > 3 ~GeV^2$,
necessary to get dilepton invariant masses well above the
vector meson region, many of them are excited and it may
not be possible to disentangle their form factors separately.
However, on the basis of Stoler's analysis \cite{Stoler},
which shows that the spacelike form factors of all resonances roughly
agree in the momentum range considered here,
we expect that at least an
average form factor can be extracted which is not too different
from the individual ones.

Our calculations thus show that indeed sizeable effects from time-like
electromagnetic form factors on the dilepton
mass spectra are to be expected in $\gamma +  N \rightarrow N + e^+e^-$
experiments.
Due to the different effects in forward
and backward scattering of the proton it is possible to obtain
information on the form factors for the nucleon as well as
for some of the resonances in the time-like sector in a momentum range
where so far no other information is available.

The authors like to thank to Y. Wada for providing us with his earlier
results and his original source code.

\newpage
\section*{Figure captions}

\bigskip
\noindent
{\bf Fig. 1} Feynman diagrams photoproduction of dileptons in the
{\it s}-channel (a), in the {\it u}-channel (b) and  the Bethe-Heitler diagrams
(c).

\bigskip
\noindent
{\bf Fig. 2} Special symmetric kinematic to reduce the influence
of the Bethe-Heitler diagrams (see ref. \cite{WALECKA}).

\bigskip
\noindent
{\bf Fig. 3}
The differential cross section for $\omega_{lab} = 1.2 ~GeV$ and
$\theta < 90^0$.
Here $\Omega_+(\theta_+,\phi_+)$ denotes the angle between the
incoming photon and the outgoing virtual photon,
$\Omega_-(\theta_-,\phi_-)$ that between the relative momentum of
the $e^+e^-$ pair and that of the outgoing virtual photon.
The calculation was performed for
the cm angles
$\theta_+=0^0$ and $\theta_-=90^0$.
The solid line gives the results of the full
calculation including the effects of VMD, while the dash-dotted
line represents only the full calculation without the VMD.
The individual contributions (including the VMD) of the
nucleon pole and the resonances are given by the dotted and the
dashed line, respectively.

\bigskip
\noindent
{\bf Fig. 4}
The differential cross section for $\omega_{lab} = 1.2 ~GeV$ and
$\theta > 90^0$. For further details see Fig. 3, but now using
$\theta_+=180^0$.

\end{document}